\documentclass[usenatbib,useAMS,fleqn]{mnras}

\usepackage{amsmath}
\usepackage{amssymb}
\usepackage{url}
\usepackage{graphicx}
\usepackage{times}
\usepackage[utf8]{inputenc}
\usepackage{xspace}
\usepackage[american]{babel}
\usepackage{color}
\usepackage{multirow}
\usepackage{float}

\def\apjl{ApJL}
\def\apjs{ApJSS}

\def\aap{A\&A}

\title[Jet-driven and jet-less NS-NS mergers]
{Jet-driven and jet-less fireballs from compact binary mergers}
\author[O. Salafia et al.]
{O.S. Salafia$^{1,2,3}$  \thanks{E--mail: om.salafia@brera.inaf.it}, G. Ghisellini$^2$ and G. Ghirlanda$^2$
  \\  \\
$^1$ Univ. di Milano Bicocca, Dip. di Fisica ``G. Occhialini'', Piazza della Scienza 3, I-20126 Milano, Italy \\
$^2$ INAF -- Osservatorio Astronomico di Brera, via E. Bianchi 46, I-23807 Merate, Italy \\
$^3$ INFN -- Sezione di Milano-Bicocca, Piazza della Scienza 3, I-20126 Milano, Italy\\
}

\begin{document}

\pagerange{\pageref{firstpage}--\pageref{lastpage}} \pubyear{2017}

\maketitle
\label{firstpage}

\begin{abstract}
During a compact binary merger involving at least one neutron star, a small fraction of the 
gravitational energy could be liberated in such a way to accelerate a small fraction 
($\sim 10^{-6}$) of the neutron star mass in an isotropic or quasi-isotropic way.
In presence of certain conditions, a pair-loaded fireball can form, which undergoes
accelerated expansion reaching relativistic velocities. As in the standard fireball scenario, 
internal energy is partly transformed into kinetic energy. At the photospheric radius, 
the internal radiation can escape, giving rise to a pulse that lasts for a time equal to the 
delay time since the merger. The subsequent interaction with the interstellar medium can then 
convert part of the remaining kinetic energy back into radiation in a weak isotropic afterglow 
at all wavelengths. This scenario does not require the presence of a jet: 
the associated isotropic prompt and afterglow emission should be visible for all NS-NS and
BH-NS mergers within 90 Mpc, independent of their inclination. The prompt emission is similar 
to that expected from an off-axis jet, either structured or much slower than usually assumed 
($\Gamma\sim 10$), or from the jet cocoon. 
The predicted afterglow emission properties can discriminate among these scenarios.
\end{abstract}


\section{Introduction}
The detection and identification of an electromagnetic (EM) counterpart to a gravitational wave (GW) event is the next
fundamental step for assessing the properties of the progenitors, refine
the distance inferred from the GW signal alone and improve our general
understanding of these phenomena. 

Black hole--black hole (BH--BH) mergers likely produce no EM signature 
\citep[even though some possibilities have been proposed, see e.g.][]{Yamazaki2016,Perna2016,Loeb2016}. 
We are left with neutron star--neutron star (NS--NS) or black hole--neutron star (BH--NS) mergers.
These are the long foreseen progenitors of Short Gamma Ray Bursts (SGRBs),
whose typical observed (isotropically equivalent) radiated energy is around $E_{\rm iso}\sim 10^{51}$ erg,
with a typical spectral peak energy $E_{\rm peak}\sim 0.1-1$ MeV \citep[e.g.][]{DAvanzo2014}.
The SGRB emission is thought to be produced by a relativistic, collimated jet
of half-opening angle $\theta_{\rm jet}\sim 5$--15 degrees \citep{Berger2014,Troja2016}.
If the jet bulk Lorentz factor $\Gamma$ is large, say $\Gamma\gtrsim 100$, 
this implies that we can detect only the small fraction $f_b=(1-\cos\theta_{\rm jet})\sim 10^{-2}-10^{-3}$ 
of SGRB jets that are observed within the half-opening angle (i.e. with a viewing angle 
$\theta_{\rm view}\le \theta_{\rm jet}$). 

If the jet is {\it structured}, with $\Gamma$ (and/or the energetics) decreasing with increasing angular
distance from the jet axis, we have somewhat better chances to see the SGRB prompt emission at large viewing angles
\citep[e.g.][]{Salafia2015,Pescalli2015,Kathirgamaraju2017}.

In order to reach relativistic velocity, the jet must excavate its way through the merger (dynamical) 
and post-merger (disk--viscosity--driven and neutrino--driven) ejecta, which may pollute
significantly the environment ahead the formation of the jet \citep{Murguia-Berthier2016,Just2015}. 
In doing so, the jet deposits part of its energy in a cocoon, which may produce a 
quasi--isotropic prompt, high--energy EM signal soon after the merger by 
photospheric release of relic thermal photons (the ``cocoon prompt emission" -- 
\citealt{Lazzati2017,Lazzati2016}) possibly followed, on a time-scale of hours, 
by UV/optical emission powered by nuclear-decay and cooling
(the ``cocoon UV emission" -- \citealt{Gottlieb2017}) and, on 
a time scale of days, by synchrotron emission 
upon interaction of the expanding cocoon with the ISM (the ``cocoon afterglow" 
-- \citealt{Lazzati2016}).
The energy the jet deposits in the cocoon is of order $\sim 10^{49}$ erg.
For the cocoon prompt emission, \citet{Lazzati2016} predict a short ($\sim 1$ s) signal 
with an energy of $\sim 10^{46}$-- $10^{47}$ ergs peaking at $\sim$ 10 keV.

The possibilities just mentioned (structured jet and energized cocoon) both
require the presence of a jet. 
In this paper we investigate the possibility to have a detectable isotropic emission in 
hard X-rays in NS-NS and BH-NS mergers {\it without} a jet.
We are guided by the fact that isolated magnetars can produce giant flares 
with $E_{\rm iso}\sim 10^{46}$ erg \citep[e.g.][]{Hurley2005,Lazzati2005} in non-catastrophic events, probably due to
some re-configuration of their magnetic field \citep{Thompson1995}.
In NS-NS and BH-NS mergers, the gravitational energy available during the last 
phase of the coalescence is more than $E_{\rm G}\sim 10^{53}$ erg, and only 
a very small fraction of this needs to be used.

To explore this possibility, we postulate that a small fraction (e.g. less than 0.1\%) 
of $E_{\rm G}$ can be used, e.g.~through magnetic field amplification and subsequent
conversion of magnetic field energy into thermal and/or kinetic energy,
during the \textbf{initial} phase of the coalescence. The magnetic field in the
NS material during the merger reaches values larger than $B\sim 10^{15}$ G \citep{Ruiz2017,Giacomazzo2014,Kiuchi2014,Zrake2013,Price2006}.
This is enough to form a fireball, that can produce an important isotropic
emission in hard X-rays, as first suggested by \citet{Zrake2013} and \citet{Giacomazzo2014}.

Various recent numerical simulations \citep[e.g.][]{Ruiz2017,Murguia-Berthier2016,Just2015} and 
theoretical works \citep[e.g.][]{Margalit2015} suggest that the conditions for launching a relativistic jet after a NS-NS or BH-NS merger are not always satisfied, thus the fraction of mergers without jets might be significant. The possible presence of a prompt, isotropic, high-energy component is thus particularly relevant for what concerns the prospects for future associations of EM counterparts to nearby GW events: \textbf{the other most promising candidate EM counterpart so far, that is the nuclear-decay-powered emission from the merger and post-merger ejecta \citep[the so-called \textit{kilonova},][]{Li1998,Metzger2016}, is predicted to emit in the Optical and Near Infrared, where no all-sky monitoring instruments are available to date.}

\section{Set up of the isotropic fireball model}

The binding energy difference between an isolated 
neutron star and a black hole of the same mass is
\begin{equation}
\Delta E_{\rm G} \sim \frac{3}{10} M_{\rm NS} c^2
\left(1-\frac{R_{\rm S}}{R_{\rm NS}}\right) \sim 5\times 10^{53}\,\rm erg
\end{equation}
where $R_{\rm S}$ is the Schwarzschild radius, $M_{\rm NS}$ and $R_{\rm NS}$ 
are the mass and radius of the NS, and the numerical value is 
for $M_{\rm NS}=1.4\,\rm{M_\odot}$ and $R_{\rm NS}= 14\,\rm km$.

This energy is of the same order as that released in a
supernova explosion. In that case, most of it goes into
neutrinos. In a compact binary merger it powers several other processes, 
such as emission of gravitational waves, dynamical ejection 
of matter and magnetic field amplification. The latter process is 
widely believed to produce a magnetic field of order $B\sim 10^{15}$--$10^{16}$ G \citep[e.g.][]{Ruiz2017,Giacomazzo2014,Kiuchi2014,Zrake2013,Price2006}, which in turn
can contain an energy of order $\sim 10^{51}\,\rm erg$ \citep{Giacomazzo2014}.

As first suggested by \cite{Zrake2013}, if just 1 percent of this energy, $E_0\sim 10^{49}\,\rm erg$, 
is converted into photons in a relatively baryon--free volume surrounding the merger 
\citep[e.g. with a mechanism similar to that responsible for giant flares in magnetars, 
see e.g.][]{Thompson1995}, a fireball initially dominated by electron-positron pairs can form and produce a short, 
high-energy transient.

From this point on, the evolution resembles that of the classical, standard {\it isotropic}
fireball of gama-ray bursts: the fireball accelerates to relativistic speed, up to  
some saturation radius $R_{\rm a}$ where the bulk kinetic energy becomes of the same order
as the initial internal energy.
Thereafter, the fireball coasts with constant velocity. At some point, the
transparency radius $R_{\rm t}$ is reached, after which the fireball
can release radiation.

The origin of the emission is still a controversial issue in the field of GRBs.
On one hand, the release of ``relic'' thermal photons (i.e. the photons
that constituted the initial source of pressure of the fireball, diluted by 
the expansion)  at the transparency (photospheric) radius is expected \citep{Meszaros2000}. 
On the other hand, in the case of GRBs the observed spectra suggest a non-thermal origin 
of the radiation, which could be the result of internal shocks \citep{Rees1994} or reconnection 
of the carried magnetic field \citep{Thompson1994} transforming part of the
kinetic energy of the fireball back into radiation. 

These processes may be present in our isotropic case as well, but 
let us first consider the thermal photospheric radiation.

The temperature of the initial blackbody can be estimated by equating the energy 
density of photons to that of the source magnetic field, i.e.\ 
$aT_0^4 \sim B^2/(8\pi)$, giving $T_0\sim 4.8\times 10^{10} B_{15}^{1/2}$ K.
The acceleration ends when the bulk Lorentz factor equals $E_0/(Mc^2)$, i.e.\ 
when \textbf{$\Gamma\sim 11\, E_{0,49}/M_{27}$ (here and in what follows we employ the usual notation $Q_x\equiv Q/10^x$ in cgs units)}.
This occurs at $R_{\rm a} = \Gamma R_0$, and there the comoving temperature 
is $T^\prime_{\rm a} = T_0(R_0/R_{\rm a})$.
Beyond $R_{\rm a}$ the temperature decreases as $T^\prime = (R /R_{\rm a})^{-2/3}$ 
and $\Gamma$ is constant. During these phases, the main radiative processes 
(pair creation and annihilation and Compton scatterings) conserve the number of 
blackbody photons.
This implies that  the total energy contained in this ``fossil'' thermal component is
\begin{equation}
E_{\rm BB} (R_{\rm t})\, \approx \, E_0\, \Gamma \, \frac{T^\prime(R_{\rm t})}{T_0}
\, =\, E_0\, \left[ \frac{R_{\rm t}}{R_{\rm a}}  \right]^{-2/3}  
\label{ebb} 
\end{equation}

The transparency radius depends on the thickness of the fireball \citep{Daigne2002}.
In our case the fireball is thin ($R_{\rm t}  \gg 2\Gamma^2 c t_{\rm inj}$)
and $R_{\rm t}$ is given by (see Eq. 15 of \citealt{Daigne2002}):
\begin{equation}
R_{\rm t}\, = \, \left[ \frac{ \sigma_{\rm T} E_0}{4\pi m_{\rm p}c^2 \Gamma} \right]^{1/2}
\, \sim 5.9\times 10^{12} \left[ \frac{ E_{0,49}}{\Gamma_1} \right]^{1/2}\,\, {\rm cm}
\end{equation}
In this case the final blackbody energy (Eq. \ref{ebb}) can be written in terms
of the two variables $E_0$ and $M$:
\begin{eqnarray}
E_{\rm BB} (R_{\rm t})\, &=& \, \frac{E_0}{M}\, \left[ \frac{ E_0 R_0}{c^2 }\right]^{2/3}
\left[ \frac{4\pi m_{\rm p}}{\sigma_{\rm T}}  \right]^{1/3} \nonumber \\
&\sim & 4\times 10^{45} \frac{E_{0,49}}{M_{27} } \, (E_{0,49} R_{0, 6})^{2/3} \, \, {\rm erg}
\end{eqnarray}
With the fiducial values of the parameters, this emission is a very small fraction
of the initial energy.
As discussed above, there can be another mechanism able to convert a larger fraction
of the bulk kinetic energy into radiation, just as in GRBs. 
Since there is no general consensus about this process, we leave it unspecified.
This ignorance is enucleated in the efficiency parameter $\eta$, such that the energy
released in this additional, non-thermal component is: 
\begin{equation}
E_{\rm \gamma} \, =\, \eta E_0
\end{equation}
With $\eta=10^{-2}\eta_{-2}$ we have $E_{\rm \gamma}= 10^{47} \eta_{-2} E_{0,49}$ erg.
Assuming a limiting sensitivity of $10^{-7}$ erg cm$^{-2}$,
such a fireball can be detected up to a distance of $\sim$90 Mpc, close to the current
horizon of LIGO/Virgo. 
This holds assuming that most of the electromagnetic radiation falls into the observed band 
of the current instruments (i.e. between 10 and 1000 keV).
This requires a radiation mechanism not only able to transform a fraction $\eta\gtrsim 10^{-2}$ of $E_0$ 
into radiation, but also able to do that in the hard X-rays.
Quasi thermal Comptonization and/or synchrotron emission are the first candidates, as
they are in ``standard'' GRBs.

\subsection{Delay and pulse duration} 

A point not always appreciated, in the GRB standard fireball theory,
is that the delay time between the initial formation and the 
arrival of the fireball at the transparency radius $R_{\rm t}$ is equal to
the duration of the pulse, if the fireball is thin (namely its width $\Delta R\ll R_{\rm t)}$.

Indeed, for a thin fireball the pulse duration is comparable to the angular time-scale
\begin{equation}
t_{\rm ang}=\frac{ R_{\rm t}}{2 c \Gamma^2} 
\sim  1\,\frac{ E^{1/2}_{0,49}}{\Gamma^{5/2}_1}\, {\rm s}
\end{equation}
The delay with respect to the initial formation, on the other hand, is given by:
\begin{equation}
\Delta t_{\rm delay}=\frac{R_{\rm t}}{c} (1-\beta) =\frac{ R_{\rm t}}{2 c \Gamma^2} 
\end{equation}
The two time scales are exactly equal.
In general, in GRBs the delay time is not measurable, 
but in the case of the detection of gravitational waves, it can be taken
as the time difference $\Delta t_{\rm delay}$ between the merger and the 
arrival of the first photons of the prompt.

\subsection{Afterglow}

After having produced the prompt emission, the fireball coasts up to the deceleration 
radius $R_{\rm dec}$, where it collects enough matter of the interstellar medium (ISM) 
to start to decelerate.
The bolometric light curve corresponding to the interaction of the fireball with the ISM 
is characterized by a rising $\propto t^2$ phase during the coasting, and a subsequent 
decay as $\Gamma$ decreases \citep{Nava2013}.
The deceleration {\it onset time} $t_{\rm on}$ is given by 
\citep[see][homogeneous case]{Nava2013}:
\begin{eqnarray}
t_{\rm on}\, &=& \, 0.48\, \left[ \frac{E_0}{n_0 m_{\rm p} c^5 \Gamma^8}  \right]^{1/3}\,  {\rm s}
\nonumber \\
\, &=& \, 6.5\times 10^4\,\left[ \frac{E_{0,49}}{n_{0, -3} \Gamma_1^8}  \right]^{1/3}\,  {\rm s}
\label{eq:tonset}
\end{eqnarray}
\textbf{This corresponds to 0.75 days. If $\Gamma=5$, $t_{\rm on}\sim 5$ days.}

\vskip 0.3 cm

\begin{figure*}
\includegraphics[width=0.8\textwidth]{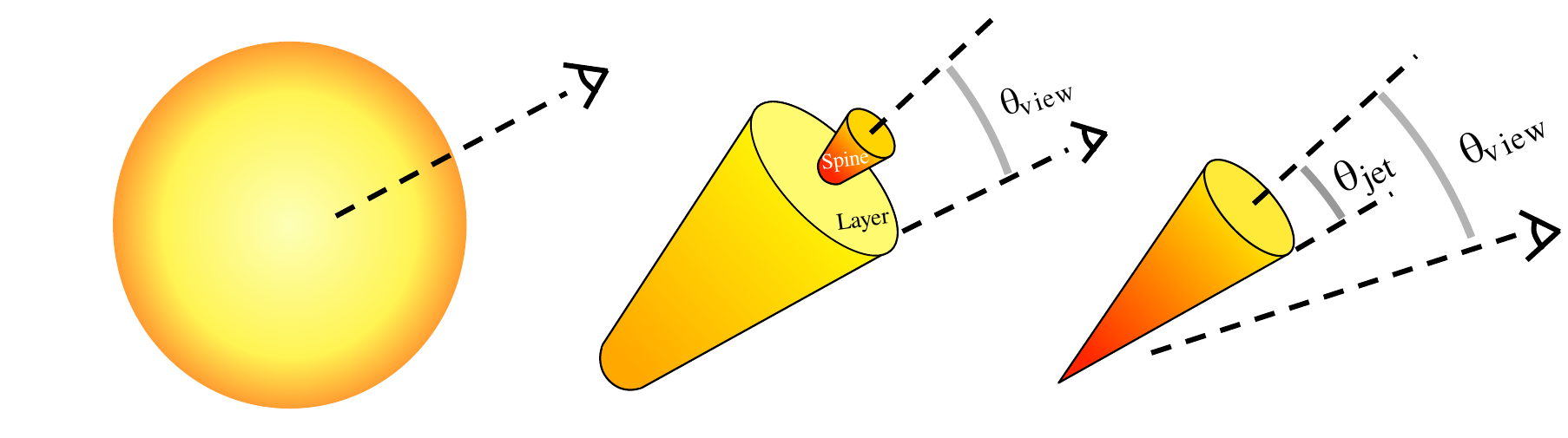}
\includegraphics[width=0.8\textwidth]{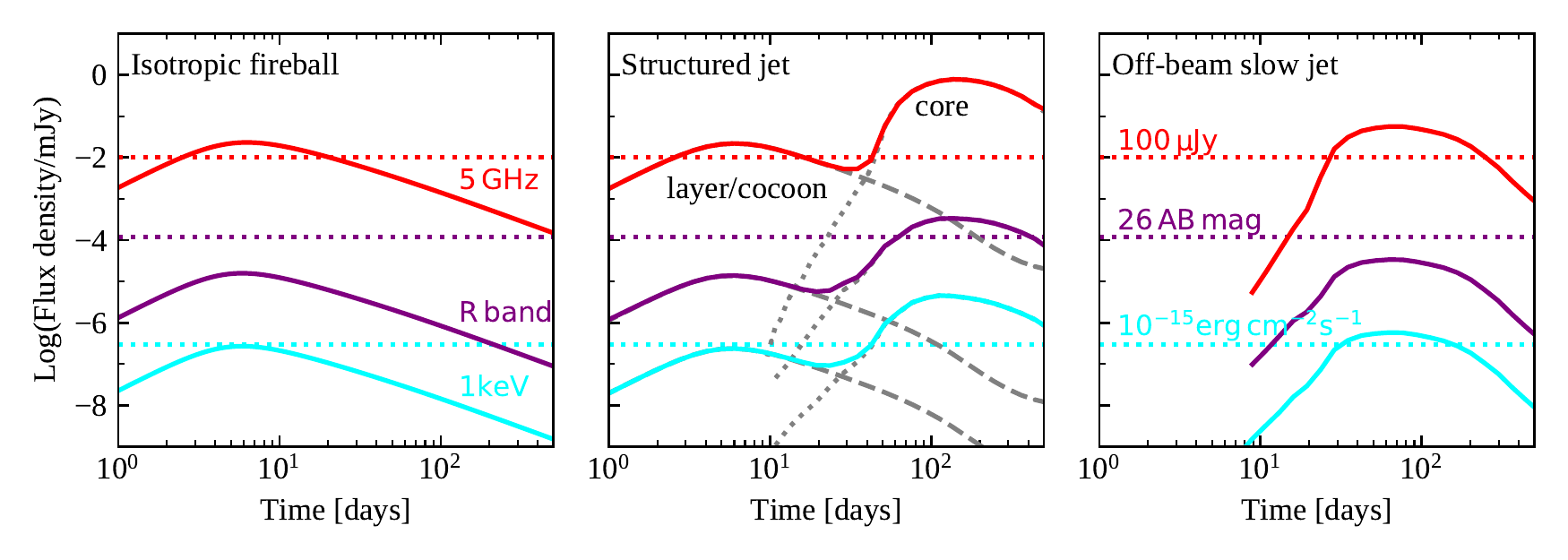}
\caption{\textbf{Indicative} afterglow lightcurves in radio (5 GHz), optical (R band) and X-ray (1 keV)
for the \textbf{three scenarios}. Parameters in Table~\ref{para}.
Left panel: isotropic fireball. 
Mid panel: the spine/layer model \textbf{(which we consider equivalent also to the jet plus cocoon scenario)}, where the spine is the assumed to have parameters typical of a standard SGRB jet.
The layer contributes to the light curve at early times, 
since $\theta_{\rm layer}=\theta_{\rm view}=30^\circ$.
Right panel: off-beam slow jet model. 
The dotted lines show representative detection limits for the three bands. The corresponding flux
values are shown on the figure. The optical and X-ray fluxes are uncertain,
since they depend on the presence of high energy electrons.
\textbf{For all models, the behaviour of the lightcurve is almost the same regardless of the observer frequency: this is due to all three frequencies being between $\nu_{\rm m}$ and $\nu_{\rm c}$ (for the chosen parameters) at all times shown on the plot, except for the early rising part of the off-beam slow jet and of the core.}
} 
\label{fig1}
\end{figure*}

\section{Comparison of the afterglow emission from the three scenarios}

It is instructive to compare the afterglow of the isotropic fireball 
with the other competing models able to explain the prompt emission of sub-energetic 
short GRBs. As mentioned in the introduction, we have the quasi-isotropic cocoon plus 
relativistic jet model by \citet{Lazzati2017,Lazzati2016} which is essentially equivalent
to the structured jet model \citep[e.g.][]{Kathirgamaraju2017}.
To this, we add what we call the off-beam slow jet, namely a rather weak and homogeneous 
jet, with moderate $\Gamma$, seen off-beam. \textbf{In order to describe the general behaviour of the afterglow in the three scenarios, we construct example lightcurves with what we think are representative parameters.}
For all models we use the same microphysical parameters (see Table~\ref{para}),
and we assume that the number density of interstellar medium (ISM) is constant
and equal for all models. \textbf{For definiteness, we locate the source at a luminosity distance $d_{\rm L}=60\,{\rm Mpc}$
which is comparable to the LIGO interferometer network range at the end of the O1 observing run \citep{Abbott2016} 
and represents a conservative estimate of the range during O2. This is also the limiting distance at which a strong
magnetar giant flare as the one described in \citet{Lazzati2005} (which we take as a prototype for our isotropic fireball scenario) can be detected by a \textit{Fermi}-like instrument,
assuming a limiting fluence of $10^{-7}\,{\rm erg\,cm^{-2}}$. We compute the lightcurves of the jetted components with the public 
code \texttt{BOXFIT} \citep{VanEerten2011}, and those of the isotropic component with \texttt{SPHEREFIT} \citep{Leventis2012}. Since neither code accounts for the dynamics before the deceleration time, we effectively correct the lightcurves by smoothly joining, at the peak time given by Eq.~\ref{eq:tonset}, a power law rising as $t^2$ at all frequencies \citep{Sari1999}}.

\subsection{Isotropic fireball}

Fig. \ref{fig1} shows (left panel) the predicted afterglow (in the radio, optical and X-rays)
of the isotropic fireball. The onset time is close to 5 days, after which the flux decays as a power law.
If a jet is not present, this is all we see.
If there is a jet with standard parameters (middle panel), it becomes visible at later times ($\sim$ 100 days), 
when it has slowed down so that $1/\Gamma \gtrsim (\theta_{\rm view}-\theta_{\rm jet})$, i.e.~when we 
start to see its border.
Its emission in the optical and X-rays depends on the uncertain presence, at late times, of
very high energy electrons.
Emission in the radio, instead, is more secure.
This late emission can be used as a diagnostic to distinguish between the jet-less, isotropic
fireball scenario and the cocoon model by \citet{Lazzati2016}, which has similar
features, but requires the presence of a jet.

\subsection{Jet plus cocoon}

Recently, a structured jet model (i.e. a jet with  both $\Gamma$ and luminosity per unit solid angle 
decreasing with the angular distance from the jet axis) has been proposed to describe 
SGRB jets when seen off-axis \citep{Kathirgamaraju2017}. \textbf{To our understanding, the mildly
relativistic layer/sheath found there is essentially the same structure as the cocoon described in \citet{Lazzati2017}.
In this work we thus consider the core-layer and the jet-plus-cocoon scenarios as equivalent.}
The middle panel of Fig.~\ref{fig1} mimics the expected afterglow from this \textbf{scenario,
obtained by summing the emission from a fast, narrow jet and that from a slower, wider jet, which represents the layer or the cocoon} (parameters in Table \ref{para}).
Since the latter has a wider half-opening angle, it is more likely seen within its beam.
The \textbf{layer afterglow peaks earlier} ($\sim 5$ days for the chosen parameters, which are the
same as for the isotropic fireball), with the spine contributing after $\sim$100 days.
As studied in \citet{rossi-polarization-2004}, in this case a rather strong linear
polarization should be present around the time when the lightcurve peaks, in contrast 
with the isotropic fireball scenario. This can be used as a diagnostic.

\subsection{Off-beam slow jet}

The energy distribution of SGRB jets may have a low energy tail,
accompanied by a corresponding low $\Gamma$ tail.
In this case a jet with $\theta_{\rm jet}\sim 10^\circ$, $E_{\rm k, iso}\sim 10^{51}$ erg
and $\Gamma\sim$5--15 could be seen also at a relatively large $\theta_{\rm view}\sim 30^\circ$.
If seen on-axis, such a jet would produce 
$E_{\rm \gamma, iso} = \eta E_{\rm k, iso} = 10^{50} \eta_{-1} E_{\rm k, iso, 51}$ erg.
These values of $E_{\rm k, iso}$ and $\Gamma$ are roughly consistent
(i.e. they are within the rather large dispersion) with the relation
shown in \citet{ghirlanda-comoving-2012} and \citet{Liang2013}.
Using Eq. 2 and Eq. 3 in \citet{Ghisellini2006} we can calculate 
the observed $E_{\rm \gamma, iso}$ for any $\theta_{\rm view}$. 
It turns out that for $\theta_{\rm view} =30^\circ$, the de-beaming
factor is $1/2500$, and then  
$E_{\rm \gamma, iso}(30^\circ)\approx 4\times 10^{46} \eta_{-1} E_{\rm k, iso, 51}$ erg,
detectable up to  $\sim$60 Mpc if the fluence limit is $10^{-7}$ erg cm$^{-2}$.
For $\theta_{\rm view}>30^\circ$ the de-beaming makes the source undetectable.
The probability to see a burst within $30^\circ$ is $P=(1-\cos30^\circ) \sim$ 0.13: small,
but not impossible. Taking into account the anisotropy of the GW emission \citep{Schutz2011},
the probability to see the jet within $30^\circ$ \textit{after the progenitor NS-NS binary has been 
detected in GW} is significantly larger\footnote{This relies upon the assumption that the 
jet is launched perpendicular to the orbital plane of the binary. This most likely holds in NS-NS mergers,
while it is less certain for BH-NS mergers, because the BH spin might cause the accretion disk plane
in the post-merger phase to be tilted with respect to the original orbital plane.}, being $\sim$ 0.38.
With the parameters listed in Table \ref{para}, we calculated the expected afterglow,
shown in the right panel of Fig. \ref{fig1}.
The peak flux corresponds approximately to the time when the Lorentz factor
$1/\Gamma\sim (\theta_{\rm view}-\theta_{\rm jet}$), namely when we start to see 
the border of the jet. 
This occurs at $t_{\rm peak}\sim$ 60 days for the parmeters shown in Tab.~\ref{para}, and 
it goes as $t_{\rm peak} \propto \left(E_{\rm k,iso}/n_0\right)^{1/3}$ for a given viewing angle,
i.e.\ it is independent from the initial Lorentz factor, due to the self-similar nature of
the deceleration \citep{blandford-blastwaves1976}.

After $t_{\rm peak}$, the flux decreases monotonically.
After the peak, the lightcurve is similar to an isotropic fireball
with the same $E_{\rm k, iso}$ and initial $\Gamma$.
However, there is an important difference: the flux of an off-beam jet should be
strongly polarized at $t_{\rm peak}$, because the observer sees only the border
of the jet \citep{rossi-polarization-2004}.

\begin{table} 
\centering
\begin{tabular}{lllll lllll lll}
\hline
\hline
Model            &$E_{\rm k, iso}$ &$\Gamma$ &$\theta_{\rm jet}$ &$\theta_{\rm view}$     \\
~                &[erg]              &         &[deg]                &[deg]                  \\
\hline 
Isotropic fireball &$10^{49}$           &5   &---  &---   \\
spine              &$10^{52}$   &100 &10   &30   \\
layer              &$10^{49}$   &5   &30   &30   \\
off--beam          &$10^{51}$           &10  &10   &30    \\
\hline
\hline 
\end{tabular}
\vskip 0.2 true cm
\caption{
Parameters used for the models shown in Fig. \ref{fig1}. For all models, we assumed
an ISM number density $n_0=10^{-3}$ cm$^{-3}$ and microphysical parameters $p=2.3$,  
$\epsilon_{\rm e}=0.1$, $\epsilon_{\rm B}=0.01$. The source is located at
a luminosity distance of 60 Mpc.
}
\label{para}
\end{table}

\section{Discussion}

Most models that associate a detectable high-energy, prompt electromagnetic 
counterpart to a gravitational wave event require the presence of a jet.
For such counterpart to be detectable at large viewing angles, the jet must be structured,
or it must be relatively slow (in order for the emission not to be too de-beamed).
It can also be a standard jet, but it must deposit part of 
its energy in a cocoon that then expands quasi-isotropically.

In this work we have explored the alternative possibility of the
liberation of a fraction of the gravitational energy of the merging objects,
instants before the merger. 
Several hints point towards the possibility that the energy required to
form a fireball might reside in the amplified magnetic field.
If a magnetized neutron star with $B\sim 10^{14}\,\rm G$ can produce $10^{46}$--$10^{47}$ erg 
in radiation in a giant flare due to a reconfiguration of its magnetic field, without
dramatic consequences on its structure, we are confident that much more
energy can be liberated during the merger. 
This is also borne out by numerical simulations, which show that magnetic field amplification
to values $B\sim 10^{15}\rm G$ or larger is a natural feature of the merger phase.
This magnetic energy may then produce an isotropic fireball, that can behave like a standard one 
even if its total energy (close to or larger than $E_0\sim 10^{49}$ erg) is much smaller than the kinetic energy
of a standard SGRB.
With this energy (and for an efficiency $\eta\sim 10^{-2}$), the prompt emission of this fireball is above the detection threshold of the {\it Fermi}/GBM or the {\it Swift}/BAT instruments if the source is within a distance of $\lesssim 90$ Mpc, which is comparable to the current binary NS range of the LIGO/Virgo detector network.\\
\indent\textbf{Several conditions are required to form a jet and they are likely not always satisfied \citep[as discussed in e.g.][]{Ruiz2017,Murguia-Berthier2016,Margalit2015,Just2015}, so that a relativistic jet is not necessarily associated to each merger. Nevertheless, our isotropic fireball scenario shows that a relatively low-luminosity gamma-ray prompt emission can be associated to a NS-NS merger even in the absence of such a jet. This implies that the observation of a high-energy prompt counterpart by itself does not necessarily point to the presence of a jet, which can instead only be revealed by late-time observation of the afterglow, possibly complemented by polarimetry.}\\ 
\indent Future detections of prompt high-energy transients associated to GW events will enable tests
of these scenarios. 
Should a jet be always detected, it would imply that the conditions for jet production
are always satisfied in the associated compact binary mergers. In turn, this has implications 
on the SGRB intrinsic rate, or equivalently on the SGRB typical half-opening angle, since it affects 
the ratio between the rate of SGRBs that we see on-axis and the total rate of NS-NS (or BH-NS)
events \citep{Ghirlanda2016}.
Until around 100 days, the afterglow of the isotropic fireball and that of the quasi-isotropic 
cocoon of \citet{Lazzati2016} should be very similar, if they start with
a comparable $E_0$ and $\Gamma$. 
What distinguishes them is mainly the presence or not of a ``standard'' jet. 
If the cocoon is much less isotropic, as seems to be the case \citep{Lazzati2017}, its afterglow may have an earlier onset, and it should show polarization at all frequencies also at early times (close to the cocoon onset time). 
The same should occur also for the off-beam slow jet, that should show polarization close to the
peak time, when the emission is dominated by the jet border.\\
\indent To conclude, we would like to stress that the possibility to have a prompt 
and an afterglow emission even without a jet has far reaching consequences
on the search for the electromagnetic counterparts to gravitational wave events,
on the physics itself of the merger, and on the typical aperture angle of the jets of SGRBs.

\section*{Acknowledgements}
We are grateful to Monica Colpi and Albino Perego for useful discussions. 

\footnotesize{
\bibliographystyle{mnras}
\bibliography{/home/omsharan/Dropbox/Bibliography/library}
}

\label{lastpage}

\end{document}